\documentclass[aps,prl,twocolumn,groupedaddress]{revtex4}

\usepackage{graphicx}
\usepackage{dcolumn}

\begin{document}


\title{Ferromagnetism and Metamagnetism in Copper-doped 
Germanium Clathrate}


\author{Yang Li}
\author{Weiping Gou}
\author{Ji Chi}
\author{Joseph H. Ross, Jr.}
\affiliation{Department of Physics, Texas A\&M University, College Station, TX 77843-4242}

\date{\today}

\begin{abstract}
Cu-doped type-I germanium 
clathrate can exhibit dilute magnetism,
including ferromagnetism, antiferromagnetism, and metamagnetic
transitions up to 90 K. 
$^{63}$Cu NMR measurements confirm that these transitions
are due to a dilute composition of magnetic
defects coupled by conduction electrons, behavior similar to that
of magnetic semiconductors.  Magnetic measurements
indicate localized magnetic moments,
attributed to clusters of magnetic ions, with competition
between ferromagnetic and antiferromagnetic exchange, and
also indications of glassy behavior in the ferromagnetic
phase.  
NMR Knight shifts and relaxation 
times show that the conduction band is metallic with a large 
Korringa ratio.  Comparison to a mean-field theory 
for the ordering behavior gives a good accounting for the ferromagnetic
transition.
\end{abstract}


\maketitle


Magnetic semiconductors have been the focus of renewed
interest, due in part to
the potential for spin injection in semiconductor 
electronics \cite{ohno98,ohno00}.  There have been a
number of recent successes in the magnetic doping 
of semiconductors, for instance GaAs can be doped out of
equilibrium with Mn using low-temperature molecular-beam 
epitaxy \cite{ohno98} to 
yield transition temperatures reported
as high as 160 K. Ferromagnetism in these materials
is due to the combined action of magnetic ions and
a polarized conduction band of itinerant electrons.
The magnetic ground state can be
complex due to the competition between locally
ferromagnetic and antiferromagnetic exchange couplings, 
as well as the positional disorder of
dopant atoms \cite{dietl01,konig00,schliemann02,zarand02}.
Thus, the nature of the ordering process is of fundamental interest.

The clathrate structure is one way to stabilize magnetic ions in group-IV
semiconductors. 
Clathrates \cite{kasper65} feature fullerene-type cages 
in a crystalline framework
enclosing electron-donating ions such as Na and Ba.  The
electron-donors can be balanced by
electron-deficient transition-metals substituted onto
the framework \cite{cordier91}, to produce materials 
ranging from semiconducting to metallic.
This is in contrast to GaAs:Mn, where Mn ions 
produce strong hole-doping. The open clathrate crystal
structure can lead to flat electron band
features, and therefore sharp density of states peaks \cite{gryko98}, 
which enhance the tendency toward
magnetism and superconductivity.  In this letter, we describe the
magnetism of a dilute magnetic Ge clathrate exhibiting competition
between ferromagnetic and antiferromagnetic interactions. 

Group-IV clathrates have a wide variety of electronic
properties.  The observations of superconductivity
\cite{kawaji95} and ``electron crystal, phonon glass'' 
behavior \cite{nolas98} in Ge and Si clathrates have sparked 
particular interest. The cages can be
vacated, leaving a new elemental form of Si \cite{gryko00}, and 
there is a possibility that these materials
may grow epitaxially on a diamond substrate \cite{munetoh01}. 
Ferromagnetism has been observed
in Mn-substituted Ge clathrate \cite{kawaguchi00}
and in clathrates with Eu ions in the cages
\cite{chakoumakos01}. Here we show that dilute Cu-doped
Ge clathrates can exhibit metamagnetism and ferromagnetism,
mediated by the conduction electrons. 

For sample preparation, ingots of Ba$_{8}$Cu$_{x}$Ge$_{46-x}$, 
were induction melted, then reacted for 3 days at $950^{\circ}$ C and 4 days at
$700^{\circ}$ C in evacuated ampoules. Cu substitutes on the 46 site/cell Ge 
framework in this case \cite{kasper65}. 
Diffraction and microprobe analysis \cite{lisubmitted} have 
shown that the type-I clathrate of approximate
composition Ba$_{8}$Cu$_{5}$Ge$_{40}$ is stabilized by a Zintl mechanism, with 
vacancies appearing spontaneously to maintain the electron count.
By changing the starting material,
the Cu content varies over a narrow range.  The samples described
here, denoted Cu2 and Cu6, had starting compositions $x$ = 2 and 6, respectively.  
Microprobe analysis showed that in the Cu6
sample, the clathrate contains 8\% additional Cu.  In addition, this
sample contains a small amount of
Ge$_{3}$Cu$_{5}$ phase intermixed with the clathrate, while
the Cu2 sample has no other Cu-containing phases \cite{lisubmitted}.

\begin{figure}
\includegraphics{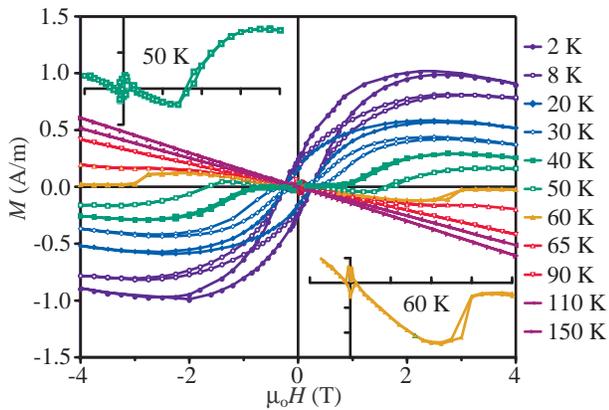}
\caption{\label{fig:fig1}Magnetization vs. field for Cu6 sample. 
The labeled temperatures have the same vertical ordering as the
curves on the right side of the figure.  Expanded views for 50 K and
60 K are inset, showing the metamagnetic response at these
temperatures.}
\end{figure}

Fig.~\ref{fig:fig1} shows magnetization ($M$) curves for the Cu6 sample,
obtained using a Quantum Design SQUID magnetometer.  
Typical ferromagnetic hysteresis is seen at 2 K, going over to a 
diamagnetic slope at high fields as the ferromagnetic response becomes 
saturated, showing the underlying diamagnetism of the lattice.
At intermediate temperatures the response is metamagnetic,
shown in the insets to Fig.~\ref{fig:fig1} as a
field-induced transition to ferromagnetic order.
The transition is first-order at 60 K, then disappears at higher temperatures.

\begin{figure}
\includegraphics{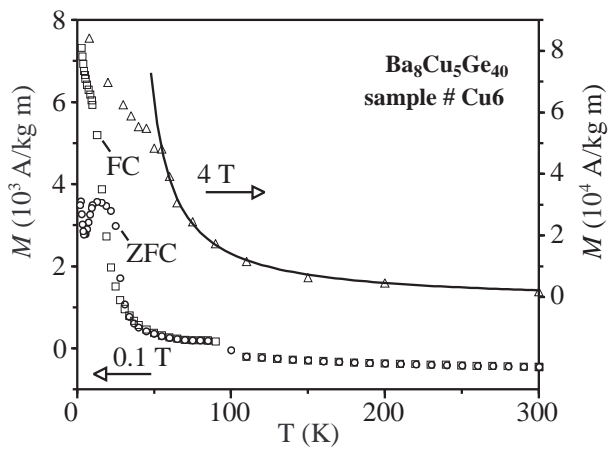}
\caption{\label{fig:fig2}Magnetization vs. temperature for
Cu-doped clathrate.  0.1 T data are
field-cooled (FC, squares) and zero-field cooled (ZFC, circles) as indicated.
Solid curve is Curie-law fit to 4 T data. }
\end{figure}

$M$ vs. $T$ has been plotted in Fig.~\ref{fig:fig2}. 
In low fields there is a feature at 90 K, associated
with an antiferromagnetic transition for the Cu6 sample,
since the sample remains diamagnetic below this 
temperature. The field-cooled (FC) curve 
has a further cusp at 12 K, associated with a
driven ferromagnetic response of the conduction electrons. 
In larger fields, the
90 K feature becomes less distinct and eventually disappears, 
as shown for 4 T applied field. A Curie law fits the 4 T data
between 55 K and 400 K (solid curve), 
with $T_{c}$ = 34 K. 

The 0.1 T zero-field-cooled (ZFC) and FC data in 
Fig.~\ref{fig:fig2} show irreversibility below 32 K. 
Furthermore, the data cross, behavior which has been
observed in re-entrant spin glass 
systems \cite{coles93,oner01}, in which a
ferromagnetic state disorders at low
temperatures due to random antiferromagnetic
bonds, and field-alignment locks some regions
into reverse-orientation, reducing the net $M$.

The Cu2 sample also exhibits low-temperature ferromagnetism. 
Spontaneous magnetization appears near 10 K, with a saturation 
moment similar to
that of Cu6.  However, no metamagnetism was observed for
the Cu2 sample, indicating the concentration-dependent nature of
this behavior.  

\begin{figure}
\includegraphics{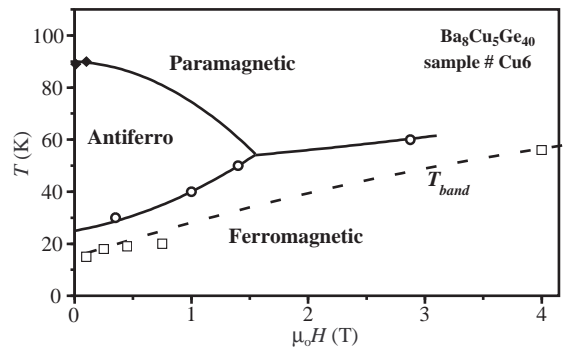}
\caption{\label{fig:fig3}Phase diagram for the metamagnetic Cu6 
sample, with an approximate rendering of the phase boundaries. 
T$_{band}$ is described in the text.}
\end{figure}

Results for Cu6 are summarized in
the phase diagram of Fig.~\ref{fig:fig3}. Circles
are metamagnetic transitions identified from
slope changes or jumps in magnetization
(Fig.~\ref{fig:fig1}).  The enclosed
phase is antiferromagnetic.  The metamagnetic 
transition changes from second-order to first order between
50 K and 60 K, as seen from the insets to 
Fig.~\ref{fig:fig1}. Thus we have plotted a
tri-critical point between these temperatures. 
The first-order curve ends in a critical point beyond which the 
paramagnetic state goes continuously into the ferromagnetic
state, shown by the unbroken Curie behavior seen in
4 T (Fig.~\ref{fig:fig2}).  
This phase diagram is quite similar to that of
NdRu$_{2}$Ge$_{2}$ \cite{gignoux95}.  
Metamagnetism is not uncommon in such
rare-earth intermetallics due to frustrated ferromagnetic
and antiferromagnetic coupling of rare earth moments, mediated
by the conduction electrons.  Here the local moments are due to
the random distribution of Cu on Ge framework sites in the clathrate.

The square symbols in Fig.~\ref{fig:fig3} identify further slope
changes in $M$ vs. $T$, for example the 4 T data departing from 
the Curie fit in Fig.~\ref{fig:fig2}).  At low fields these 
appear below 
the metamagnetic line.  As shown below, this feature is 
consistent with the saturation of the
band electrons, thus the label $T_{band}$.

Since $M$ approaches the lattice diamagnetic slope
at high fields (Fig.~\ref{fig:fig1}), we measured the 400 K response
in order to establish this slope, and thus determine the 
low-temperature saturation 
moment, $M_{s}$.  With an applied field of 7 T, we obtained
$M_{s} = 0.085 \mu_{B}$ per cell in the Cu6 sample.  
Since $M_{s} = gJ\mu_{B}n$, and the
Curie law is $\chi = n\mu_{o}\mu_{B}^{2}p_{eff}^{2}/3k(T-T_{c})$, 
the measurements may be used
to evaluate $n$, the spin density, and $p_{eff}=g\sqrt{J(J+1)}$,
the local moment. In our case $J$ is unknown, however
assuming $J$ = 1/2, the 4 T Curie fit
(Fig.~\ref{fig:fig2}) gives $n$ = 0.023 per 
cell and $p_{eff} = 6.2\mu_{B}$.
Alternatively, $J$ = 5/2 gives $n$ = 0.011 per 
cell and $p_{eff} = 9.2\mu_{B}$, in either case a dilute set of moments

In type-I Ge clathrates, transition metal atoms generally
substitute on the 6$c$ crystallographic site, 6 per unit
cell among the 46 framework sites \cite{cordier91}. Using NMR
as a local probe of the Cu sites, we find that the main line
consists of non-magnetic Cu in a high-symmetry environment, consistent
with occupation of the 6$c$ site. The moments of 6.2 or 9.2 $\mu_{B}$
are too large to be single ions, and are presumed associated with 
composite clusters of neighboring atoms, possibly centered upon
wrong-site Cu atoms. Competition between antiferromagnetic 
superexchange and ferromagnetic double exchange is well known
for magnetic semiconductors \cite{dietl01,methfessel68}, and this 
can explain the metamagnetic behavior observed here:  competing 
internal interactions in the clusters, in addition to the
RKKY interaction between clusters, can cause a switching
in the magnetic alignment.  

\begin{figure}
\includegraphics{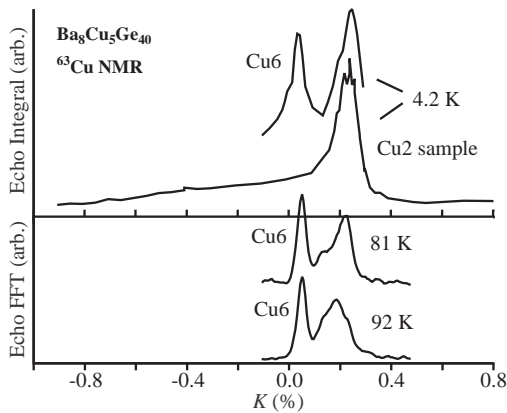}
\caption{\label{fig:fig4}$^{63}$Cu NMR spectra for the
two samples. 
4.2 K data measured via
echo integration; lower two traces measured via small tip-angle
echo FFT.
Data offset vertically for clarity.}
\end{figure}

$^{63}$Cu NMR confirms that the dilute magnetism is intrinsic to the clathrate.
Data were obtained in a fixed applied field of 9 T. Above 4 K, data were 
obtained by FFT of small-tip-angle spin echoes, allowing
the frequency range to be extended.
Fig.~\ref{fig:fig4} shows NMR lineshapes,
on a Knight shift ($K$) scale based on an
aqueous $^{27}$AlCl$_{3}$ reference and published gyromagnetic
ratios \cite{carter77}.  The peak near $K$ = 0.2\% is
common to both samples, while Cu6 has an additional lower-$K$ peak,
changing little with temperature, due to the presence of Ge$_{3}$Cu$_{5}$
in that sample.   A low-$K$ tail was also seen for both samples, 
as shown in the 4.2 K Cu2 data.  This tail
can be attributed to nuclei in direct contact with magnetic 
sites, while the main peak is more characteristic of a non-magnetic
metal: Cu metal has $K$ = 0.239 \%
\cite{carter77}, while the typical range for Cu compounds
and nonmagnetic intermetallics is 0 -- 0.2 \%.

We attributed the low-$K$ NMR peak to Ge$_{3}$Cu$_{5}$ 
since it appears only in the Cu6 sample.  
(Ge$_{3}$Cu$_{5}$ is apparently
stabilized by the clathrate; attempts to produce  
this phase alone gave instead a mixture of Ge and
orthorhombic GeCu$_{3}$, with a $T$-independent 
spectrum different from those of Fig.~\ref{fig:fig4}.)
The higher-$K$ peak has the symmetry expected for the
nearly-tetrahedral clathrate lattice. The
6$c$ site has four identical nearest neighbors,
and X-ray diffraction of the Cu clathrate shows the
bond angles for this site to be within 0.5$^{\circ}$ of
perfect tetrahedral symmetry \cite{lisubmitted}.
The other framework sites also approximate a tetrahedral
bonding configuration, but with somewhat more distorted 
bond angles and crystallographically inequivalent neighbors.
For the 6$c$ site, we expect small electric field gradients
and small electric quadrupole splittings in $^{63}$Cu NMR,
while lower symmetry gives typically a
central transition flanked by quadrupole-split satellites.  
The matrix elements determining the NMR pulse angles
differ by a factor of two for these cases \cite{kanert71}.
Our investigation of the pulse response showed that the 
lower-$K$ Cu6 line is a central transition, corresponding
to a lower-symmetry environment in the unknown Ge$_{3}$Cu$_{5}$
material.  The other Cu6 line, and the 
corresponding line in Cu2, has little or no quadrupole splitting,
characteristic of a high-symmetry environment, consistent
with its assignment to Cu occupying the 6$c$ site of the
clathrate framework. 

\begin{figure}
\includegraphics{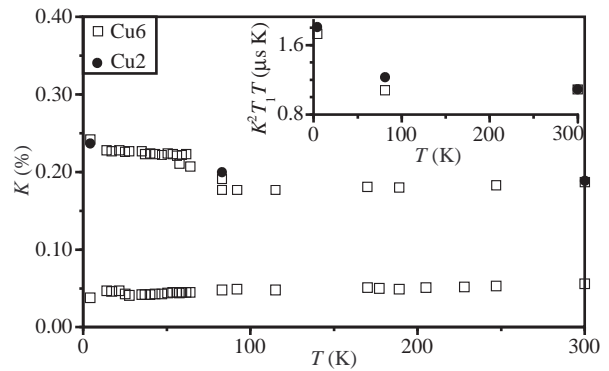}
\caption{\label{fig:fig5}$^{63}$Cu NMR peak Knight shifts vs. temperature.
Inset shows the $K^{2}T_{1}T$ product.}
\end{figure}

Fig. ~\ref{fig:fig5} shows $K$ vs. $T$,
with points representing lineshape maxima.
For the clathrate 6$c$-line in both samples, $K$ increases at low
$T$.  This increase is most prominent for the
range 60 -- 90 K in Cu6, and we also
observed a warming/cooling hysteresis between 77 K 
and 90 K.
Extrapolating the curve $T_{band}$ (Fig. ~\ref{fig:fig3})
to the 9 T field of the NMR spectrometer,
we see that 
changes in $K$ occur
over the range where the band moment
is expected to be changing most rapidly. Thus, the 6$c$ 
resonance exhibits changes that correspond to changes
in the magnetism, which we
confirm to be intrinsic to the clathrate.

The NMR spin-lattice relaxation time, $T_{1}$, 
follows a Korringa law in both samples, indicating the 
presence of a metallic conduction band.  
$T_{1}$ was obtained by inversion-recovery,
with a fit to a single exponential recovery curve for the
6$c$ site, appropriate
for cases with no quadrupole splitting \cite{kanert71}. 
The product $K^{2}T_{1}T$ is shown
in the inset of Fig.~\ref{fig:fig5}. For both samples, 
$K^{2}T_{1}T$ is nearly constant above 90 K, characteristic
of relaxation by a metallic band \cite{carter77}. The increase 
in $K^{2}T_{1}T$
at low temperatures is due almost entirely to changes in $K$.
$K^{2}T_{1}T$ values are smaller than that expected for the 
bare nucleus, $3.8 \mu$s K, which can come about if the
Knight shift is a difference of terms, with the negative
contribution due to core polarization, for which the
hyperfine coupling is negative \cite{carter77}. 
This indicates
a contribution of Cu $d$-electrons to the conduction band.
However, the fact that the $T_{1}$ change is much smaller than
that of $K$ indicates that many of the carriers are unaffected by
the transition, probably due to multiple bands crossing the
Fermi surface \cite{gryko98}.

The low-$T$ increase in $K$ is consistent with a
band polarized in the opposite sense to the local moments,
assuming that the shift is dominated by core polarization,
which has negative hyperfine coupling for Cu.
Using the coupling constant $H^{cp} = -$12.5 T \cite{carter77}, the 
low-temperature change in $K$,
+0.073 \% for Cu6, corresponds to a saturated conduction-band 
magnetization $M=-0.0032 \mu_{B}$ per cell. In this calculation
we assumed the band magnetization to be distributed equally among the
46 framework sites. If we also approximate by assuming
that this magnetization is linear in the defect magnetization,
we can use the mean-field transition temperature calculated
for
conduction-band-coupled localized moments \cite{jungwirth02}: 
\begin{equation}
\ k_{B}T_{c} = \frac{np_{eff}J_{int}^{2}\chi}{4\pi(g^{*}\mu_{B})^{2}}. \label{eq:one}
\end{equation}
Here $J_{int}$ is the coefficient of a delta-function interaction
between the local moments and the band, $\chi$ is the 
band susceptibility, and $g^{*}$ its g-factor.  
Assuming the conduction
electrons are uniformly polarized by the effective Zeeman
field corresponding to $J_{int}$, 
we can write $M = \chi nJgJ_{int}/(\mu_{o}g^{*}\mu_{B})$
for the band magnetization which we evaluated using the
Knight shift.  Using 
$J$ = 5/2, $n$ = 1.1\%, and $p_{eff} = 9.2$ as extracted
above, $g^{*}$=2, and $T_{c}$=12 K, we
obtain
$J_{int}$=0.27 eV nm$^{3}$ and $\chi=1.4 \times 10^{-5}$
(SI dimensionless).
These are plausible values; $J_{pd}$ = 0.055 eV nm$^{3}$ 
was obtained for Mn in 
GaAs \cite{jungwirth02}, while the $\chi$ we obtain is about 
one-fifth that of copper metal \cite{carter77}. 

Since the susceptibility is in the metallic range,
and the magnetic defects well separated, one might
expect that spin glass ordering would result from the
oscillatory RKKY interaction. However, the clathrate structure can
lead to large band masses and sharp electron density of states
features \cite{gryko98}, which may enlarge the 
oscillation period simply by the decrease of the Fermi
wavevector, or possibly by enhancing the response for $k$ = 0
in the sense of a paramagnon system \cite{moriya85}. These
features could make related clathrate systems useful for
future spin-related applications.

Thus, we have shown that the Cu-doped Ge clathrate
can exhibit a rich magnetic
structure, including an antiferromagnetic phase which
features a metamagnetic transition to the ground-state 
ferromagnetic configuration.  NMR measurements confirmed 
this behavior to be due to a dilute set of 
magnetic clusters, coupled via the conduction-band electrons.  

\begin{acknowledgments}
This work was supported by the Robert A. Welch Foundation, 
Grant No. A-1526, and by Texas A\&M University through the
Telecommunications and Informatics Task Force.
\end{acknowledgments}

\bibliography{clathrates}

\end{document}